# THIRD INTEGER RESONANCE SLOW EXTRACTION USING RFKO AT HIGH SPACE CHARGE.*

V.Nagaslaev#, J. Amundson, J.Johnstone C.S.Park, S.Werkema,
FNAL, Batavia, IL 60510, U.S.A.

*Abstract*

A proposal to search for direct μ→e conversion at Fermilab requires slow, resonant extraction of an intense proton beam. Large space charge forces will present challenges, partly due to the substantial betatron tune spread. The main challenges will be maintaining a uniform spill profile and moderate losses at the septum. We propose to use "radio frequency knockout" (RFKO) for fine tuning the extraction. Strategies for the use of the RFKO method will be discussed here in the context of the Mu2e experiment. The feasibility of this method has been demonstrated in simulations.

## INTRODUCTION

The Mu2e experiment is proposed at FNAL to search for rare neutrinoless decays of a muon to an electron in the Coulomb field of the atomic nucleus [1]. The design sensitivity of this experiment is unprecedented $6\times10^{-17}$, and requires a very strong suppression of the background. The Fermilab Debuncher ring will provide a slow spill with a pulsed longitudinal structure and very clean gaps between bunches for this purpose. A single short bunch per beam turn, slowly extracted from the Debuncher, gives an interval between pulses that is equal to the Debuncher revolution period of 1.69μs. This is almost ideal for the experiment's requirements. Additional suppression in the gaps between pulses is provided by an external extinction system, that removes out-of-time beam at level of $10^{-10}$.

The 8 GeV kinetic energy proton beam originates from the FNAL Booster and is sent to the Accumulator via the Recycler. Three batches of 53MHz Booster beam are momentum stacked and then rebunched into h=4, 2.5MHz rf buckets. Beam bunches are then sequentially transferred, one at a time, to the Debuncher and slowly extracted to the Mu2e target over a time interval of 160ms.

## RESONANT EXTRACTION

Third order resonant slow extraction is the base line design for the Mu2e project, because it promises potentially better extraction efficiency, which is crucial for 24kW beam operation. The details of the extraction scheme have been described elsewhere [2], where the idea of using RFKO for spill rate control is introduced and its feasibility is shown. Computer simulations of third-integer resonant extraction have been performed using the ORBIT code developed at ORNL [3]. We also have used this code for investigating the RFKO details presented here. The detector conditions of the Mu2e experiment set strict requirements on the uniformity of the spill. The main challenge of satisfying the spill uniformity requirement is the large space charge tune spread. In the current design conditions SC tune shift is expected to be 0.012-0.015.

Figure 1 shows the distributions of particle tune versus the horizontal action at the onset of the resonance (a), and at the point when the machine tune reaches the resonance point 29/3.

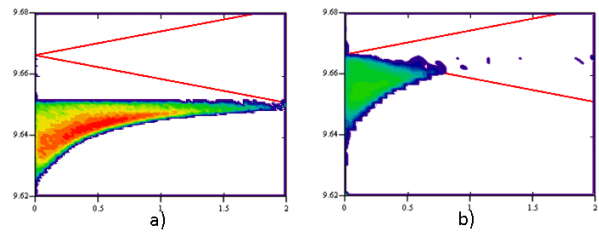

Figure 1. Tune distributions vs. horizontal action at the onset of the resonance and at exact resonance.

Red lines show the 2/3 resonance extraction area boundaries due to a sextupole field that is imposed to create a 3$^{rd}$ order separatrix. When the machine tune reaches 29/3, a substantial part of the beam remains and is far from the resonance. The tune ramp is stopped at this point, extraction continues, and the tune spread shrinks. This helps the extraction rate, but takes a long time. Clearly, it is a very challenging task to adjust tune ramp so that the extraction rate is uniform, especially since the rate at which quadrupole field corrections can be made is very limited.

## RF KNOCK-OUT

We propose another way to assist extraction in the situation close to Figure 1b. Using an RF horizontal kick, one can effectively heat the beam transversely and therefore assist the transition of particles through the separatrix into the unstable region. With proper mixing, it also assists in depopulating the low amplitude part of the distribution, therefore helping to create a more uniform tune distribution.

This technique is known as RF knock-out (RFKO). It has been already used for slow extraction purposes in medical applications [4], although the primary use in these applications is to turn off/on the beam extraction.

___________________________________________
* Work supported by DOE under contract No. DE-AC02-07CH11359
#vnagasl@fnal.gov

Our goal is to use RFKO as a tool for the fine control of the spill rate.

RFKO allows us to continue extraction in the presence of the strong space charge while keeping the machine tune close to the resonance. In this case particles are extracted on the resonance; therefore the step size is maximized.

It will be shown below that a sufficient transverse E-field may be provided by a regular damper kicker. Frequency modulation around a single betatron sideband is required to sweep the excitation frequency within certain limits. Two modulation schemes have been considered: a linear sweep (chirp) and random noise within a certain frequency interval ("colored noise").

One important feature should be noted about the linear chirp modulation. Because of the periodic nature of the modulation, the actual frequency content of the signal would be a series of narrow lines. In order to destroy phase correlations causing this effect, one can randomly shift the phase after each sweep. The resulting frequency spectrum is shown in Figure 2 along with the corresponding colored noise spectrum. The spectrum has noticeable tails and a fine structure that depends on the period of the chirp.

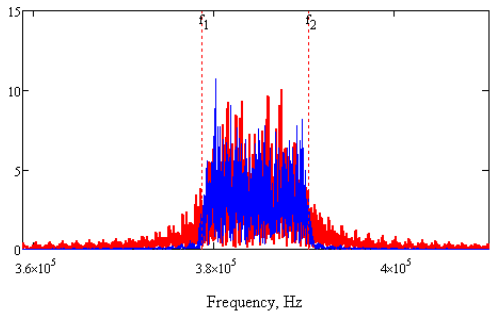

Figure 2. Frequency spectrum of the linear chirp with randomized phase (red) and colored noise (blue). Dotted markers correspond to the definition of the FM range

A subset of ORBIT simulations [2,3] has been set up to optimize the efficiency of the RFKO beam heating. Tracking simulations have been performed for two different types of the beam tune spread: chromatic tune spread and space charge (SC) tune spread. Tracking does include the motion of the bunched beam in the longitudinal RF field, however it is important to note here that for the purposes of fine spill rate control we need a sufficiently fast beam response and we are considering a time scales below one millisecond. This is well below the synchrotron period, and therefore, the beam behaviour is not much different from that of coasting beam. RFKO kick sequences were generated according to the range of power specs of commercially available amplifiers.

Figure 3 shows the beam chromatic tune distribution (red, left scale) and the rms emittance growth rate (right scale) for noise (blue) and linear chirp (green) frequency modulation versus frequency (in units of revolution frequency). The width of each bin for last two curves corresponds to the width of the FM, which is equal to 0.004.

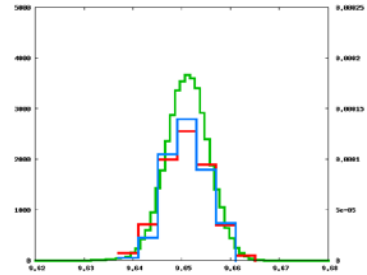

Figure 3. Beam chromatic tune distribution for $\xi_x=2$ (green) and emittance growth rate for two schemes of FM: colored noise (blue) and linear chirp (red).

As expected, the heating efficiency is proportional to the particle density within the FM band. The two types of modulation yield similar performance with small advantage for colored noise at the centre of the distribution. For the FM centered in the middle of the distribution, heating efficiency (rms emittance growth rate) versus (inverse) FM width is shown in Figure 4:

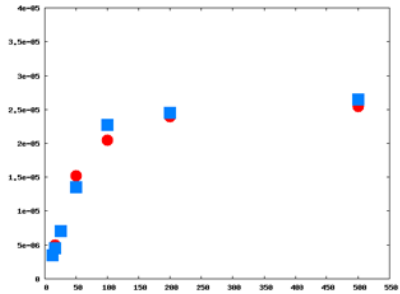

Figure 4. Heating efficiency versus inverse FM bandwidth with noise modulation (blue) and linear chirp (red). Bandwidth is used in units of revolution frequency.

For large FM BW the growth rate is proportional to the fraction of time when kicker frequency is within the beam distribution, therefore it's inversely proportional to the BW (left part of the plot). When BW is narrower than the beam spread, its effectiveness improves, but the fraction of excited particles reduces, so the net effect is a very slow growth. For practical reasons optimal FM bandwidth is equal to the beam tune spread. Keeping the BW equal to the tune distribution FWHM, we plot dependence of the growth rate on the beam tune spread in Figure 5.

As expected, the growth rate is inversely proportional to the beam tune spread (and FM BW). This can be interpreted as the time when each particle is affected by the RFKO force within certain vicinity of its betatron frequency reduces as the tune span increases. Therefore it is desirable to have a narrow tune spread and use narrow FM BW. However, running into very narrow spreads and widths presents certain problems with both: fluctuations of the force become large and damping the dipole motion becomes too slow.

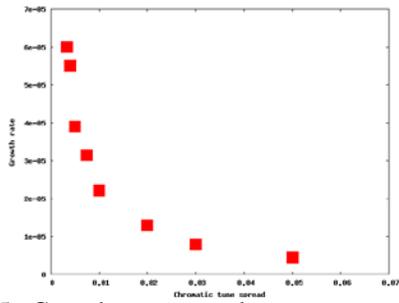

Figure 5. Growth rate versus beam tune spread. Noise FM only.

All above considerations are very different if the tune spread is induced by the space charge. The first difference is that the coherent betatron oscillations of the beam as a whole are not affected by the internal SC forces. Therefore, in order to excite dipole oscillations, the external force should be applied at the bare lattice tune frequency. Figure 6 presents the frequency scan around the tune space with addition of a small chromaticity. Again, the green trace shows the tune distribution of the beam, the blue and red curves show the rms emittance growth rates for colored noise and linear chirp modulation correspondingly. The bin width for last two curves represent the width of the FM.

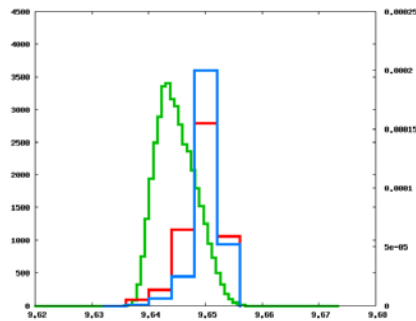

Figure 6. Frequency dependence of the heating efficiency (right scale) on the centre frequency of RFKO excitation with colored noise (blue) and chirp modulation (red). Green curve shows the tune distribution in the beam (left scale).

Apparently, the optimum heating efficiency corresponds to an RFKO excitation frequency at the machine tune (Q=0.650). Its spread corresponds to the chromatic part of the tune spread.

The second principal difference between the SC tune spread case and the chromatic tune spread case is less obvious. At zero chromaticity coherent dipole oscillations are completely decoupled from the incoherent SC motion and therefore they never decohere, and indeed, simulations show no emittance growing. For purposes of resonant extraction, pure dipole oscillations excited by RFKO would be able to move the beam out of the separatrix, however, this would result in a significant difference in intensity between consecutive micro-bunches in the spill. This is highly undesirable for the experiment. Mixing and emittance dilution must prevent the growth of the dipole amplitude, so a certain level of chromaticity need to be introduced to give a chromatic tune spread on top of the SC tune spread. This is why it is present in Figure 6. This, in turn, sets the lower limit on the FM modulation bandwidth and hence, the growth rate. There is another reality factor that must be taken into account in determining the bandwidth. There must be substantial overlap between the chromatic tune spread and the FM bandwidth, and this depends on the accuracy of RFKO following the machine tune ramping.

For large SC, overlap of chromatic tune spread and the SC tune spread is small; therefore the dependence of growth rate on the SC can be expected to be weak. Figure 7 shows rms emittance growth rate versus the SC beam tune spread. This is done for noise modulation only. For each run horizontal chromaticity was the same and equal to 1, which corresponds to 95% chromatic tune spread approximately 0.007. The dependence is not as fast as that in Figure 5, however, it is still noticeable. More RFKO power will be required for higher beam intensities.

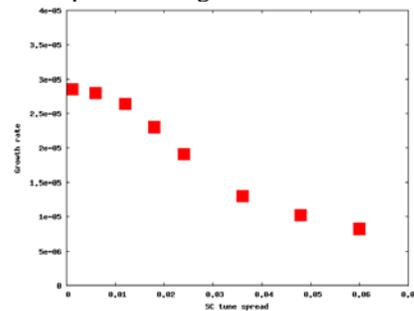

Figure 7. Growth rate dependency on the SC tune shift

## SUMMARY

Tracking simulations show that optimal RFKO frequency modulation in case of the SC beam tune spread is substantially different from that of chromatic tune spread. FM should be centred at the bare betatron frequency rather than in the middle of the tune spread, and additional chromaticity must be added in order to facilitate rapid dilution of the dipole oscillations. Heating efficiency slowly decreases with the SC growing.

Colored noise modulation (random signal within a given bandwidth) appears to be the most effective way of modulation, however its advantage over linear modulation is not very large, so practical reasons may prevail in the final choice between the two.